\documentclass[prl,twocolumn,superscriptaddress,showpacs]{revtex4}
\usepackage{graphicx,amsmath,amssymb,bm}

\renewcommand{\epsilon}{\varepsilon}

\newcommand{\be}{\begin{equation}}
\newcommand{\ee}{\end{equation}}
\newcommand{\vlowk}{V_{{\rm low}\,k}}

\newcommand{\fmi}{\, \text{fm}^{-1}}

\newcommand{\la}{\langle}
\newcommand{\ra}{\rangle}

\begin{document}

\title{In-Medium Similarity Renormalization Group for Nuclei}

\author{K. Tsukiyama}
\email[E-mail:~]{tsuki@nt.phys.s.u-tokyo.ac.jp}
\affiliation{Department of Physics, University of Tokyo, Hongo, Tokyo,
113-0033, Japan}

\author{S. K.\ Bogner}
\email[E-mail:~]{bogner@nscl.msu.edu}
\affiliation{National Superconducting Cyclotron Laboratory
and Department of Physics and Astronomy, Michigan State University,
East Lansing, Michigan 48844, USA}

\author{A. Schwenk}
\email[E-mail:~]{schwenk@physik.tu-darmstadt.de}
\affiliation{ExtreMe Matter Institute EMMI, GSI Helmholtzzentrum f\"ur
Schwerionenforschung GmbH, 64291 Darmstadt, Germany}
\affiliation{Institut f\"ur Kernphysik, Technische Universit\"at
Darmstadt, 64289 Darmstadt, Germany}
\affiliation{TRIUMF, 4004 Wesbrook Mall, Vancouver, British Columbia, V6T 2A3, Canada}

\begin{abstract}
We present a new {\it ab initio} method that uses similarity renormalization
group (SRG) techniques to continuously diagonalize nuclear many-body
Hamiltonians. In contrast with applications of the SRG to two- and
three-nucleon interactions in free space, we perform the SRG evolution
``in medium'' directly in the $A$-body system of interest. The
in-medium approach has the advantage that one can approximately evolve
$3,...,A$-body operators using only two-body machinery based on
normal-ordering techniques. The method is nonperturbative and can be
tailored to problems ranging from the diagonalization of
closed-shell nuclei to the construction of effective valence-shell Hamiltonians and operators. We present first results for the
ground-state energies of $^4$He, $^{16}$O and $^{40}$Ca, which have
accuracies comparable to coupled-cluster calculations.
\end{abstract}

\pacs{21.30.Fe, 21.60.De, 21.10.-k, 21.60.Cs}

\maketitle

Great progress has been made in ab-initio nuclear structure over the
past decade, where it is now possible to calculate properties of light
nuclei up to about carbon~\cite{GFMC,NCSM} and low-lying states of
medium-mass nuclei near closed shells~\cite{CC}. A key challenge in
nuclear physics is to extend this ab-initio frontier to larger and
open-shell systems. This requires methods that can handle the strong
coupling between low and high momenta in nuclear forces used in these
calculations.

In recent years, new approaches to nuclear forces based on
renormalization group (RG) ideas have been developed that decouple
high-momentum degrees of freedom by lowering the resolution (or a
cutoff) scale in nuclear forces to typical nuclear structure
momentum scales~\cite{vlowkreview}. Such RG-evolved potentials, known
generically as ``low-momentum interactions,'' greatly simplify the
nuclear many-body problem and enhance the convergence of structure and
reaction calculations, while the freedom to vary the resolution scale
provides a powerful tool to assess theoretical uncertainties due to
truncations in the Hamiltonian and from many-body
approximations~\cite{vlowkreview,bognerSRG,jurgenson3Nsrg,nucmatt}.

One path to decouple high-momentum degrees of freedom is the
similarity renormalization group (SRG), which was introduced
independently by Glazek and Wilson~\cite{glazek1993} and
Wegner~\cite{wegner1994}. The SRG consists of a continuous sequence of
unitary transformations that suppress off-diagonal matrix elements,
driving the Hamiltonian towards a band- or block-diagonal
form. Writing the unitarily transformed Hamiltonian as
\be
 H(s) = U(s) H U^\dagger(s) \equiv H^{\rm d}(s)+H^{\rm od}(s) \,,
\label{eq:ham_uni_trans}
\ee
where $H^{\rm d}(s)$ and $H^{\rm od}(s)$ are the appropriately defined
``diagonal'' and ``off-diagonal'' parts of the Hamiltonian, the
evolution with the flow parameter $s$ is given by
\be
\frac{dH(s)}{ds} = [\eta(s),H(s)] \,.
\label{eq:fs_srg_flow_eqn}
\ee
Here $\eta(s) \equiv [dU(s)/ds] \, U^{\dagger}(s)$ is the
anti-Hermitian generator of the transformation. The choice of the
generator first suggested by Wegner,
\be
 \eta(s) = [H^{\rm d}(s),H(s)] = [H^{\rm d}(s),H^{\rm od}(s)] \,,
\label{eq:eta_weg4}
\ee
guarantees that the off-diagonal coupling of $H^{\rm od}$ is driven
exponentially to zero with increasing $s$~\cite{wegner1994}. Through
different choices for $H^{\rm d}$ and $H^{\rm od}$, one can tailor the
SRG evolution to transform the initial Hamiltonian to a form that is
most convenient for a particular problem~\cite{kehrein2006,white2002}.
It is this flexibility, together with the fact that one never
explicitly constructs and applies the unitary transformation $U(s)$
[rather it is implemented implicitly through the integration of
Eq.~(\ref{eq:fs_srg_flow_eqn})] that makes the SRG a powerful
alternative to conventional effective interaction methods such as
Lee-Suzuki similarity transformations~\cite{vlowkreview}.

To date, the SRG applications to nuclear forces have been carried out
in free space to construct ``soft'' nucleon-nucleon (NN) and
three-nucleon (3N) interactions to be used as input in ab-initio
calculations~\cite{vlowkreview,SRGUCOM}. While
the free-space evolution is convenient, as it does not have to be
performed for each different nucleus or nuclear matter density, it is
necessary to handle 3N (and possibly higher-body) interactions to be
able to lower the cutoff significantly and maintain approximate
cutoff independence of $A \geqslant 3$ observables. The SRG
evolution of 3N operators represents a significant technical challenge
that has only recently been solved in a convenient
basis~\cite{jurgenson3Nsrg}.

\begin{figure}[t]
\begin{center}
\includegraphics[width=8.0cm,clip=]{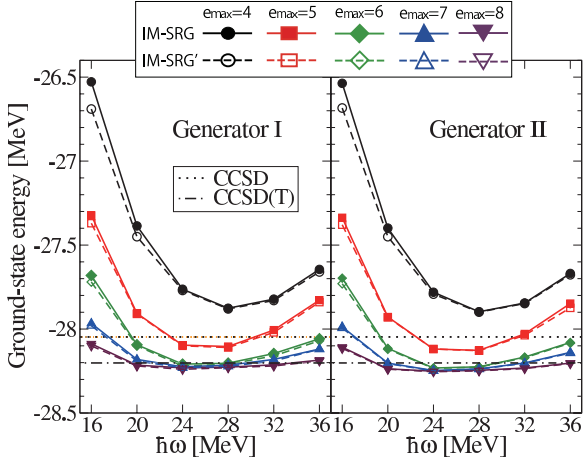}
\end{center}
\vspace*{-4mm}
\caption{Convergence of the in-medium SRG results at the
normal-ordered two-body level, IM-SRG(2), for $^{4}$He using the
generators $\eta^{\rm I}$ (left) and $\eta^{\rm II}$ (right
panel). The filled (open) symbols correspond to solving
Eqs.~(\ref{eq:e0_flow})--(\ref{eq:2b_flow}) with the underlined terms
omitted (included). The ground-state energy $E_0(\infty)$ 
is given as a function
of the harmonic oscillator parameter $\hbar \omega$ with increasing
single-particle space $e_{\rm max}\equiv\max(2n+l)$. The initial NN
interaction is a free-space SRG-evolved potential with $\lambda=2.0
\fmi$ from the N$^3$LO potential of Ref.~\cite{entem2003n3lo}. For
comparison we show the coupled-cluster CCSD and CCSD(T) energies in
the $e_{\rm max}=8$ space (calculated at their $\hbar \omega$ minimum).
\label{fig:he4_srg2.00}}
\vspace*{-2mm}
\end{figure}

An interesting alternative is to perform the SRG evolution directly in
the $A$-body system of
interest~\cite{wegner1994,kehrein2006,white2002}. Unlike the
free-space evolution, the in-medium SRG (IM-SRG) has the appealing feature that
one can approximately evolve $3,...,A$-body operators using only
two-body machinery. The key to this simplification is the use of
normal-ordering with respect to a finite-density reference
state. Starting from a general second-quantized Hamiltonian with two-
and three-body interactions, all operators can be normal ordered with
respect to a finite-density Fermi vacuum $|\Phi\rangle$ (e.g., the
Hartree-Fock ground state), as opposed to the zero-particle
vacuum. Wick's theorem can then be used to exactly write $H$ as
\begin{align}
 H &= E_0 + \sum_{ij} f_{ij} \, \{a_i^\dagger a_j\}
+\frac{1}{2!^2}\sum_{ijkl} \Gamma_{ijkl} \, \{a_i^\dagger a_j^\dagger
a_l a_k\} \nonumber \\
&+\frac{1}{3!^2}\sum_{ijklmn} W_{ijklmn} \,
\{a_i^\dagger a_j^\dagger a_k^\dagger a_n a_m a_l\} \,,
\label{eq:normal-ordered_hamiltonian}
\end{align}
where the normal-ordered strings of creation and annihilation
operators obey $\langle\Phi|\{a^{\dagger}_i \cdots a_j\}|\Phi\rangle 
= 0$, and the normal-ordered 0-, 1-, 2-, and 3-body terms are given by
\begin{align}
 E_0 &= \la \Phi | H | \Phi \ra = \sum_{i} T_{ii} \, n_i
+\frac{1}{2} \sum_{ij} V^{(2)}_{ijij} \, n_i \, n_j \nonumber \\
&+ \frac{1}{6} \sum_{ijk} V^{(3)}_{ijkijk} \, n_i \, n_j \, n_k \,,
\label{eq:n-ordered_e0_ini} \\
f_{ij} &= T_{ij} + \sum_{k} V^{(2)}_{ikjk} \, n_k
+ \frac{1}{2} \sum_{kl} V^{(3)}_{ikljkl} \, n_k \, n_l \,, 
\label{eq:n-ordered_g_ini} \\
\Gamma_{ijkl} &= V^{(2)}_{ijkl} + \frac{1}{4} \sum_{m} V^{(3)}_{ijmklm}
\, n_m \,, \label{eq:n-ordered_gamma_ini} \\
W_{ijklmn} &= V^{(3)}_{ijklmn}.
\label{eq:n-ordered_w_ini}
\end{align}
Here, the initial $n$-body interactions are denoted by $V^{(n)}$, and
$n_i=\theta(\epsilon_{\rm F}-\epsilon_i)$ are occupation numbers in
the reference state $|\Phi \ra$, with Fermi energy $\epsilon_{\rm
F}$. It is evident from
Eqs.~(\ref{eq:n-ordered_e0_ini})--(\ref{eq:n-ordered_gamma_ini}) that
the normal-ordered terms, $E_0$, $f$ and $\Gamma$, include contributions
from the three-body interaction $V^{(3)}$ through sums over the
occupied single-particle states in the reference state
$|\Phi\rangle$. Therefore, truncating the in-medium SRG equations to
normal-ordered two-body operators, which we denote by IM-SRG(2),
will approximately evolve induced three- and higher-body interactions
through the nucleus-dependent 0-, 1-, and 2-body terms. As a preview,
we refer to Fig.~\ref{fig:he4_srg2.00} with the very promising
convergence of the $^4$He ground-state energy, which is comparable to
coupled-cluster results.

Using Wick's theorem to evaluate Eq.~(\ref{eq:fs_srg_flow_eqn}) with
$H(s) = E_0(s) + f(s)+ \Gamma(s)$ and $\eta = \eta^{(1)} + \eta^{(2)}$
truncated to normal-ordered two-body operators, one obtains the
coupled IM-SRG(2) flow equations (with $\bar{n}_i \equiv 1-n_i$):
\begin{align}
&\frac{dE_0}{ds} = \underline{\sum_{ij} \eta^{(1)}_{ij} f_{ji} \, (n_i-n_j)}
+\frac{1}{2} \sum_{ijkl} \eta^{(2)}_{ijkl} \Gamma_{klij} \,
n_i n_j \bar{n}_k \bar{n}_l ,
\label{eq:e0_flow} \\
&\frac{df_{12}}{ds} = \sum_{i} \Bigl[ \eta^{(1)}_{1i} f_{i2} 
+ (1 \leftrightarrow 2) \Bigr] \nonumber \\
&+\underline{ \sum_{ij} (n_i-n_j) (\eta^{(1)}_{ij} \, \Gamma_{j1i2}
- f_{ij} \, \eta^{(2)}_{j1i2}) }\nonumber \\
&+ \frac{1}{2} \sum_{ijk} \Bigl[ \eta^{(2)}_{k1ij} \Gamma_{ijk2} 
(n_i n_j \bar{n}_k + \bar{n}_i \bar{n}_j n_k) 
+ (1 \leftrightarrow 2) \Bigr] , \label{eq:1b_flow} \\[2mm]
&\frac{d\Gamma_{1234}}{ds} = \sum_{i} \Bigl[ (\underline{\eta^{(1)}_{1i} \,
\Gamma_{i234}} - f_{1i} \, \eta^{(2)}_{i234}) - (1 \leftrightarrow 2) \Bigr]
\nonumber \\
&- \sum_{i} \Bigl[ (\underline{\eta^{(1)}_{i3} \, \Gamma_{12i4}} - f_{i3} \,
\eta^{(2)}_{12i4}) - (3 \leftrightarrow 4) \Bigr]
\nonumber \\
&+ \frac{1}{2} \sum_{ij} \Bigl[ \eta^{(2)}_{12ij} \, \Gamma_{ij34}
(1-n_i-n_j)
+ (1,2 \leftrightarrow 3,4) \Bigr] \nonumber \\
&- \sum_{ij} (n_i - n_j) \Bigl[ (\eta^{(2)}_{j2i4} \Gamma_{i1j3}
- \eta^{(2)}_{i1j3} \Gamma_{j2i4}) - (1 \leftrightarrow 2) \Bigr] \,.
\label{eq:2b_flow}
\end{align}
The IM-SRG(2) equations exhibit important similarities to the CCSD
approximation of coupled-cluster theory. For instance, the commutator
form of the flow equations gives a fully connected structure in which
$H(s)$ has at least one contraction with $\eta$. Therefore, there are
no unlinked diagrams and the flow equations are size extensive.
Combined with the $\mathcal{O}(N^6)$ scaling with the number of
single-particle orbitals, this makes the method well suited for
calculations of medium-mass nuclei. The IM-SRG is intrinsically
nonperturbative, where the flow equations,
Eqs.~(\ref{eq:e0_flow})--(\ref{eq:2b_flow}), build up nonperturbative
physics via the interference between the particle-particle and the two
particle-hole channels for $\Gamma$ and between the
two-particle--one-hole and two-hole--one-particle channels
for~$f$. The perturbative analysis reveals that the IM-SRG(2) energy
is third-order exact (as is the CCSD approximation) and that $f$ and
$\Gamma$ are second-order exact~\cite{imsrglong}. It also implies that
for calculations with harder interactions, the underlined terms in
Eqs.~(\ref{eq:e0_flow})--(\ref{eq:2b_flow}) should be excluded because
they produce higher-order contributions (with alternating signs) to
$E_0$ that are also generated by the inclusion of higher-body
normal-ordered interactions, $\eta^{(3)}$ and $W$, corresponding to
simultaneous $3p3h$ excitations. Because such triples excitations can
be sizable for hard potentials, the underlined terms in
Eqs.~(\ref{eq:e0_flow})--(\ref{eq:2b_flow}) should be omitted to better
preserve the partial cancellations that would occur against the
$[\eta^{(3)},W]$ contributions. This is consistent with the
observation in Fig.~\ref{fig:he4_srg2.00} that for soft potentials
our results are insensitive to the inclusion of these terms. Therefore
we define the IM-SRG(2) truncation without these terms for consistency.

\begin{figure}[t]
\begin{center}
\includegraphics[width=8.0cm,clip=]{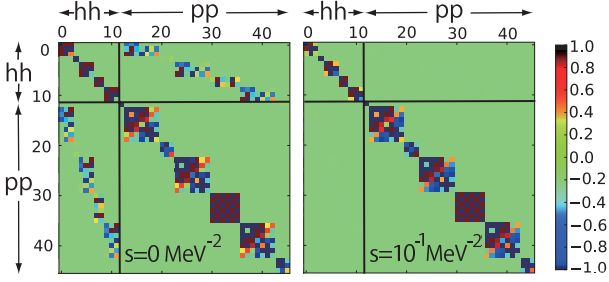}
\end{center}
\vspace*{-2mm}
\caption{In-medium SRG evolution of normal-ordered two-body matrix
elements $\Gamma_{ijkl}$ connecting hole-hole (hh) and
particle-particle (pp) states for $^{16}$O starting from a
smooth-cutoff $\vlowk$ with $\Lambda = 1.8 \fmi$. The color scale
is in MeV, and initial and $s$-evolved results are shown.
The axes label two-body $jj$-coupled states
$|(n_a,l_a,j_a,t_{z_a}),(n_b,l_b,j_b,t_{z_b});J=0 \ra$. The
$\Gamma_{ijkl}$ where $ijkl=ppph$ or $hhhp$, which are not driven to
zero with the current generator, are not shown.}
\label{fig:tbme_evolution}
\vspace*{-2mm}
\end{figure}

In this initial study, we restrict our attention to the ground states
of doubly-magic nuclei and define $H^{\rm od}(s) = f^{\rm od}(s) +
\Gamma^{\rm od}(s)$, with
\begin{align}
f^{\rm od}(s) &= \sum_{ph} f_{ph}(s) \, \{a_p^\dagger a_h\} +
\text{H.c.} \,, \\
\Gamma^{\rm od}(s) &= \sum_{pp'hh'} \Gamma_{pp'hh'}(s) \, 
\{ a_{p}^\dagger a_{p'}^\dagger a_{h'} a_{h}\} + \text{H.c.} \,,
\end{align}
where $p,p'$ and $h,h'$ denote unoccupied (particle) and occupied
(hole) Hartree-Fock
orbitals, respectively. We consider two different cases for the
generator $\eta$. First, we take the Wegner choice 
$\eta^{\rm I}(s) = [H^{\rm d}(s),H^{\rm od}(s)]$. Second,
we follow White~\cite{white2002} and define
\begin{align}
\eta^{\rm II} &= \sum_{ph} \frac{f_{ph} \, \{a^\dagger_p a_h\}}{f_{p}-f_{h}
-\Gamma_{phph}} - \text{H.c.} \nonumber \\
&+ \sum_{pp'hh'} \frac{\Gamma_{pp'hh'} \, \{a^{\dagger}_p
a^{\dagger}_{p'} a_{h'} a_h\}}{f_{p}+f_{p'}-f_{h}-f_{h'} + A_{pp'hh'}} 
- \text{H.c.} \,,
\label{eq:gen_whi4}
\end{align}
where $A_{pp'hh'} = \Gamma_{pp'pp'} + \Gamma_{hh'hh'}
- \Gamma_{phph} - \Gamma_{p'h'p'h'} - \Gamma_{ph'ph'} - \Gamma_{p'hp'h}$
and $f_p\equiv f_{pp}$ (the $s$ dependence is
suppressed for simplicity). Both generators suppress off-diagonal
($1p1h$ and $2p2h$) couplings and drive
the Hamiltonian towards diagonal form,
\be
H(\infty) = E_0(\infty) + f^{\rm d}(\infty) + \Gamma^{\rm d}(\infty) \,,
\label{eq:evolved_H}
\ee
but White's generator ($\eta^{\rm II}$) is significantly more
efficient, because the flow equations are less stiff in this case and the
evaluation of $\eta$ at each step is significantly faster. The evolved
Hamiltonians using $\eta^{\rm I}$ and $\eta^{\rm II}$ are unitarily
equivalent if no truncations are made. Any differences in energy
eigenvalues therefore provide a measure of the truncation error
resulting from neglected three- and higher-body normal-ordered terms
in our calculations.

\begin{figure}[t]
\begin{center}
\includegraphics[width=7.5cm,clip=]{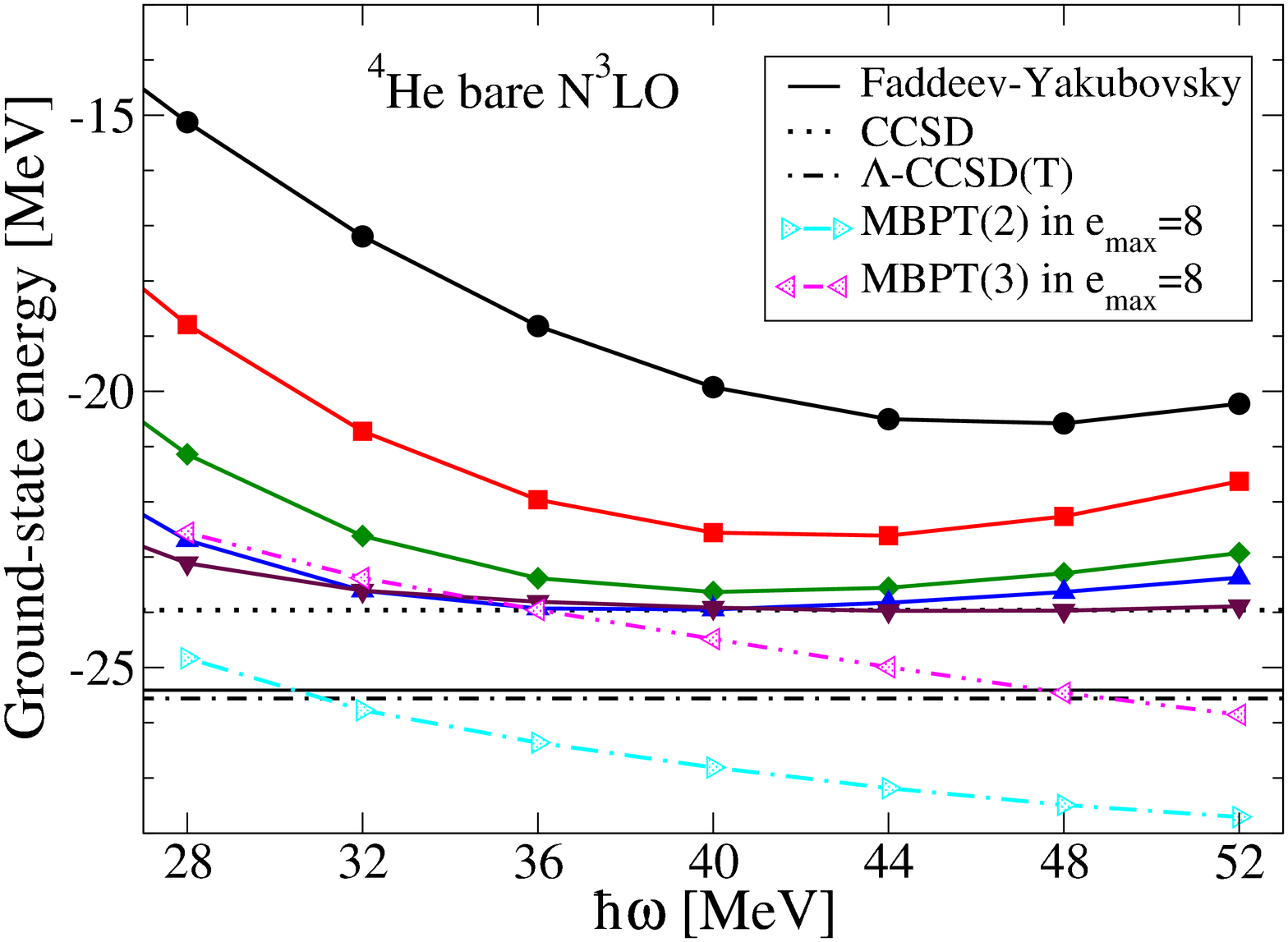}
\end{center}
\vspace*{-4mm}
\caption{Convergence of the IM-SRG(2) energy $E_0(\infty)$ for $^{4}$He
using the generator $\eta^{\rm II}$ and starting from the ``bare'' N$^3$LO
potential of Ref.~\cite{entem2003n3lo}. The notation is the same as
in Fig.~\ref{fig:he4_srg2.00}. The converged IM-SRG(2) energy agrees
well with the CCSD result (the coupled-cluster energies are taken
from Ref.~\cite{CC}), while second- and third-order many-body
perturbation theory, MBPT(2) and MBPT(3), clearly break down.
\label{fig:he4_baren3lo}}
\vspace*{-2mm}
\end{figure}

At the end of the flow, the reference state becomes the ground 
state of $H(\infty)$, with fully interacting ground-state energy
$E_0(\infty)$, and $|\Phi \ra$ decouples from the rest of the
Hilbert space ($1p1h, 2p2h, \ldots, ApAh$ sectors),
\be
QH(\infty)P = 0 \quad \text{and} \quad PH(\infty)Q = 0 \,,
\label{eq:decoupling}
\ee
where $P=| \Phi \ra \la \Phi |$ and $Q=1-P$. This decoupling
follows from the observation that all other normal-ordered couplings
annihilate the reference state, $[f^{\rm d}(s)+\Gamma^{\rm d}(s)]
|\Phi\rangle = 0$. Combined with $f^{\rm od}(\infty)$ and $\Gamma^{\rm
od}(\infty)$ being driven to zero, this implies the block-diagonal
structure of Eq.~(\ref{eq:decoupling}). The IM-SRG is very flexible
and alternative choices of $H^{\rm od}$ (and $\eta$) can be used to
target excited states, single-particle properties, and to construct
effective valence shell-model Hamiltonians and operators for
open-shell systems~\cite{kehrein2006,white2002}.

\begin{figure}[t]
\begin{center}
\includegraphics[width=8.25cm,clip=]{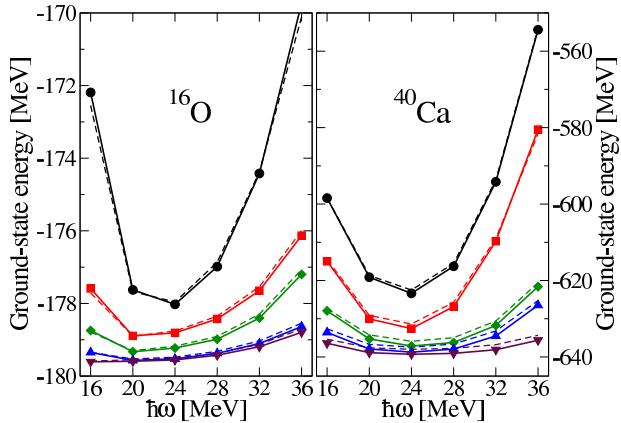}
\end{center}
\vspace*{-4mm}
\caption{Convergence of the IM-SRG(2) energy $E_0(\infty)$ for
$^{16}$O (left) and $^{40}$Ca (right panel) using the generator
$\eta^{\rm II}$ (solid symbols) and in comparison to CCSD results
(dashed lines). The notation is the same as in
Fig.~\ref{fig:he4_srg2.00}. The initial $V_{\rm NN}$ is a
smooth-cutoff $\vlowk$ with $\Lambda = 1.8 \fmi$ for $^{16}$O and a
free-space SRG potential with $\lambda = 1.8 \fmi$ for $^{40}$Ca,
both evolved from the N$^3$LO potential of
Ref.~\cite{entem2003n3lo}.\label{fig:o16_ca40_srg1.80}}
\vspace*{-2mm}
\end{figure}

Figure~\ref{fig:he4_srg2.00} shows the IM-SRG(2) ground-state energy
$E_0(\infty)$ for $^{4}$He calculated in increasing spaces defined by
the single-particle $e_{\rm max}\equiv \max(2n+l)$. For all cases the
flow equations, Eqs.~(\ref{eq:e0_flow})--(\ref{eq:2b_flow}), were
solved in a $jj$-coupled basis. The $\eta^{\rm I}$ and $\eta^{\rm II}$
results agree within $20 \, {\rm keV}$, which suggests the truncation
to normal-ordered two-body interactions is a controlled approximation.
This is consistent with Fermi system arguments for interparticle
interactions where a finite-density reference state is close to the
interacting ground state~\cite{3body}. In addition, the IM-SRG(2)
$e_{\rm max}=8$ energy is essentially converged and within $20 \, {\rm
keV}$ of the exact NCSM diagonalization~\cite{jurgenson3Nsrg}, and
in good agreement with the coupled-cluster CCSD(T) energies (based on
the code of Ref.~\cite{CC}). We stress that the agreement is obtained
at the normal-ordered two-body level without including residual
three-body interactions.

The suppression of $H^{\rm od}(s)$ is illustrated in
Fig.~\ref{fig:tbme_evolution}, which shows the $\eta^{\rm
I}$-evolution of normal-ordered two-body matrix elements
$\Gamma_{ijkl}$. As expected, the off-diagonal couplings ($ijkl=pphh$
or $hhpp$) are rapidly driven to zero. An important practical
consequence is that many-body approximations become more effective
under the SRG evolution before complete decoupling has been reached.

Figure~\ref{fig:he4_baren3lo} shows the IM-SRG(2) results for $^4$He
starting from a ``bare'' N$^3$LO potential, which is a harder initial
interaction. The ground-state energy clearly converges to a value
close to the CCSD result. The failure of many-body perturbation theory
in this case verifies that the IM-SRG is an intrinsically
nonperturbative method.

Finally, we apply the IM-SRG to calculate the ground-state energies of
$^{16}$O and $^{40}$Ca in Fig.~\ref{fig:o16_ca40_srg1.80}. As for the
$^{4}$He results of Fig.~\ref{fig:he4_baren3lo}, the calculations are
well converged and have accuracies that closely track the CCSD
energies. As discussed above, the IM-SRG(2) includes some simultaneous
$3p3h$ excitations for $E_0(s)$ that partially cancel against
contributions that would arise if normal-ordered three-body operators
were kept in the flow equations. This motivated excluding the
underlined terms in Eqs.~(\ref{eq:e0_flow})--(\ref{eq:2b_flow}). The
omitted terms are negligible for soft interactions, as shown in
Fig.~\ref{fig:he4_srg2.00}, but they become larger for hard
interactions such as the ``bare'' N$^3$LO potential used here, and
thus require a consistent treatment either by omitting them in the
IM-SRG(2) equations, or by including normal-ordered three-body
operators in the flow equations. In the former case, we find here an
accuracy that is comparable to CCSD calculations.

In summary, we have shown that the in-medium SRG is a promising method
for ab-initio calculations of light and medium-mass nuclei. The use
of normal ordering allowed us to evolve the dominant induced
$3,...,A$-body interactions using only two-body machinery. We have
presented first IM-SRG(2) results for the ground-state energies of
closed-shell nuclei, which were in very good agreement with CC
calculations. Work is in progress to include 3N forces and to
study effective valence shell-model Hamiltonians and operators for
open-shell systems.

\begin{acknowledgments}
We thank T.\ Duguet, R.\ Furnstahl, G.\ Hagen, T.\ Papenbrock, R.\
Perry and T.\ Otsuka for discussions. This work was supported in
part by the JSPS, the U.S.~Department of Energy UNEDF SciDAC
Collaboration under Contract No.~DEFC02-07ER41457, the U.S.~National
Science Foundation under Grant No.~PHY-0758125, by NSERC, the
Helmholtz Alliance Program of the Helmholtz Association, contract
HA216/EMMI ``Extremes of Density and Temperature: Cosmic Matter in
the Laboratory'' and the DFG through Grant SFB 634.
\end{acknowledgments}


\vspace*{-2mm}

\begin{thebibliography}{99}
\bibitem{GFMC} S.\ C.\ Pieper, Riv.\ Nuovo Cim.\ {\bf 031}, 709 (2008).

\bibitem{NCSM} P.\ Navr\'{a}til {\it et al.}, J.\ Phys.\ G {\bf 36},
083101 (2009).

\bibitem{CC} G.\ Hagen {\it et al.}, Phys.\ Rev.\ C {\bf 82}, 034330
(2010).

\bibitem{vlowkreview} S.\ K.\ Bogner, R.\ J.\ Furnstahl and A.\ Schwenk,
Prog.\ Part.\ Nucl.\ Phys.\ {\bf 65}, 94 (2010).

\bibitem{bognerSRG} S.\ K.\ Bogner, R.\ J.\ Furnstahl and R.\ J.\ Perry,
Phys.\ Rev.\ C {\bf 75}, 061001 (2007).

\bibitem{jurgenson3Nsrg} E.\ D.\ Jurgenson, P.\ Navratil and 
R.\ J.\ Furnstahl, Phys.\ Rev.\ Lett.\ {\bf 103}, 082501 (2009).

\bibitem{nucmatt} K.\ Hebeler {\it et al.}, Phys.\ Rev.\ C in press,
arXiv:1012.3381.

\bibitem{glazek1993} S.\ D.\ Glazek and K.\ G.\ Wilson, Phys.\ Rev.\ D
{\bf 48}, 5863 (1993).

\bibitem{wegner1994} F.\ Wegner, Ann.\ Phys.\ (Leipzig) {\bf 506}, 77 (1994).

\bibitem{kehrein2006} S.\ Kehrein, {\it The Flow Equation Approach to 
Many-Particle Systems} (Springer, Berlin, 2006).

\bibitem{white2002} S.\ R.\ White, J.\ Chem.\ Phys. {\bf 117}, 7472 (2002).

\bibitem{SRGUCOM} R.\ Roth, S.\ Reinhardt and H.\ Hergert, Phys.\ Rev.\
C {\bf 77}, 064003 (2008).

\bibitem{entem2003n3lo} D.\ R.\ Entem and R.\ Machleidt, Phys.\ Rev.\ C
{\bf 68}, 041001 (2003).

\bibitem{imsrglong} K.\ Tsukiyama, S.\ K.\ Bogner and A.\ Schwenk,
in prep.

\bibitem{3body} B.\ Friman and A.\ Schwenk, arXiv:1101.4858.
\end{thebibliography}
\end{document}